\documentclass[twocolumn,aps,prl,english,twocolumn,epsfig,groupedaddress]{revtex4}
\usepackage{graphicx}
\usepackage[latin1]{inputenc}
\begin{document}
\title{\bf Comment on "Magnetic Percolation and the Phase Diagram of the disordered RKKY model'  }

\author{Richard ~Bouzerar$^{1}$\footnote[4]{email: richard.bouzerar@u-picardie.fr} Georges ~Bouzerar$^{2}$\footnote[6]{email: georges.bouzerar@grenoble.cnrs.fr} and Timothy ~Ziman$^{3}$\footnote[5]{and CNRS, email: ziman@ill.fr} 
}

\affiliation{$^{1}$Universit\'e  de Picardie Jules Verne, 33 rue Saint-Leu, 80039 Amiens Cedex 01\\ 
$^{2}$Laboratoire Louis N\'eel 25 avenue des Martyrs, CNRS, B.P. 166 38042 Grenoble Cedex 09
France.\\   
$^{3}$Institut Laue Langevin B.P. 156 38042 Grenoble 
France.\\
}            
\date{\today}

\begin{abstract}
\end{abstract}
\parbox{14cm}{\rm}
\medskip

\pacs{PACS numbers: 75.30.Et 77.80.Bh 71.10.-w}
\maketitle

\section{}
The letter by Priour and Sarma \cite{PriourSarma2006} only confirms, by Monte Carlo
methods, results obtained previously by us within a semi-analytical approach \cite{Richard}: the fully  self-consistent local
 Random Phase Approximation(SC-LRPA)\cite{Bouzerar2}. Although very similar to their letter, we regret that our 
work is  not quoted. However, the publication of their letter is interesting for several reasons. 
First, it confirms the correctness of the phase diagram of the dilute RKKY model: Figure 4 of \cite{Richard} and Fig.1 of 
\cite{PriourSarma2006} are in essence identical. 
Additionally, as demonstrated in \cite{Richard,Bouzerar2} and contrary to many published statements, 
even those of the authors of the letter \cite{DasSarma1,DasSarma2,Ohno,Prospects}, the dilute RKKY model
is {\it not} a viable model for dilute magnetic semiconductors(DMS)
because of  frustrating effects which are not seen in the simplest mean field theory\cite{Ohno}
or VCA approximations\cite{DasSarma1,DasSarma2}. Thirdly, it clearly supports the crucial points, made by us in several papers \cite{Bouzerar2,Richard}, 
that the problem of these oversimplified 
theories is not only due to an inappropriate treament of the Heisenberg Hamiltonian 
(Mean Field Virtual Crystal Approximation) but also due to the invalidity 
of the RKKY form of magnetic couplings.
Finally, on a more theoretical level, it demonstrates
the accuracy of the SC-LRPA, both in the region of stable ferromagnetism, 
which is not surprising given the success of the calculations in 
the comparison with experimental results in diluted magnetic semi-conductors (and Monte Carlo results using
the same {\it ab initio} couplings\cite{Josef},
but even to accurately predict instabilities due to frustration effects. Given the speed of computation  
(at least 3 orders of magnitude faster) this confirms,
contrary to statements sometimes made, the superiority of the semi-analytical method. 
\par
Where the letter \cite{PriourSarma2006} falls short, is in giving a resolution
to the puzzle: if RKKY  {\it cannot} explain the ferromagnetism
of DMS for physical values of carrier density, what is missing from the model
approach? Are the model approaches of no use to  understand ferromagnetism in diluted magnetic 
semiconductors?  In our earlier paper\cite{Richard}, by examining the issue
heuristically  starting from  two impurities, and in a more recent work\cite{Jpd},  
in terms of the dilute non perturbative $J_{pd}$  model, we provide an answer. The physical mechanism
that is missing from the perturbative-based RKKY picture, and which is present
in ab-initio approaches, is that when treated non-perturbatively, the $J_{pd}$
interactions can introduce a resonant nature to the conduction band. This  gives 
a ferromagnetic bias to the magnetic interactions and stabilizes ferromagnetism.
This is seen clearly in Figure 3 of \cite{Jpd} where we demonstrate that the ferromagnetism persists
for much higher carrier densities and at higher temperatures than is possible in the RKKY model.

\end{document}